\documentclass[aps,pre, preprint]{revtex4-1}
\usepackage{amsthm}
\usepackage{subfigure}
\usepackage{graphicx}
\usepackage{epsfig}
\usepackage{epstopdf}
\usepackage{color}
\usepackage{caption}
\usepackage{amsmath}
\usepackage{amssymb}
\usepackage{hyperref}
\usepackage{soul}
\usepackage{wasysym}
\usepackage{ulem}
\theoremstyle{definition}

\begin{document}

\title{Influence of viscosity contrast on buoyantly unstable miscible fluids in porous media}
\author{Satyajit Pramanik, Tapan Kumar Hota, Manoranjan Mishra}
\affiliation{Department of Mathematics, Indian Institute of Technology Ropar, 140001 Rupnagar, India}

\begin{abstract}
The influence of viscosity contrast on buoyantly unstable miscible fluids in a porous medium is investigated through a linear stability analysis (LSA) as well as direct numerical simulations (DNS). The linear stability method implemented in this paper is based on an initial value approach, which helps to capture the onset of instability more accurately than the quasi-steady state analysis. In the absence of displacement, we show that viscosity contrast delays the onset of instability in buoyantly unstable miscible fluids. Further, it is observed that suitably choosing the viscosity contrast and injection velocity a gravitationally unstable miscible interface can be stabilised completely. Through LSA we draw a phase diagram, which shows three distinct stability regions in a parameter space spanned by the displacement velocity and the viscosity contrast. DNS are performed corresponding to parameters from each regime and the results obtained are in accordance with the linear stability results. Moreover, the conversion from a dimensionless formulation to the other and its essence to compare between two different type of flow problems associated with each dimensionless formulation are discussed. 
\end{abstract}

\maketitle

\section{Introduction}\label{sec:Intro}
Understanding hydrodynamic instabilities and mixing of miscible fluids in porous media is an active area of research, related to several industrial and environmental processes, such as oil recovery \citep{Homsy1987}, CO$_2$ sequestration \citep{Huppert2014}, groundwater contamination \citep{Brian2008}, chromatography \citep{Guiochon2008}, to name a few. Global warming attributed to anthropogenic emission of greenhouse gases is one of the major challenges facing mankind. Carbon capture and storage (or CO$_2$ sequestration) in underground aquifers is detected as a promising mean to restrict unwanted rise of greenhouse gas level in the atmosphere. Geological sequestration of disposal in deep saline aquifers offers several key assets of miscible fingering instabilities driven by both viscous and buoyancy forces. At a subsurface saline aquifer site the injected super-critical lighter CO$_2$ rises up and accumulates under an impervious rock, followed by dissolution of CO$_2$ into the underlying brine. An unstably stratified diffusive interface of heavier CO$_2$ dissolved brine is formed above the pure brine and transition to natural convection in the form of unstable sinking plumes is featured in time. 

Several theoretical \citep{Huppert2014, Loodts2014, Daniel2014} and experimental \citep{Backhaus2011, Loodts2014} studies are conducted to understand the convective instabilities devoted to characterising optimal storage sites that ensure CO$_2$ does not leak into the environment. In such convective flows, viscosity variation, albeit small, at the diffusive interface influences the onset of instability and hence plays significant role in characterising the storage sites. Despite having enormous importance in real life applications this problem remains poorly explored. Manickam and Homsy \cite{Manickam1995} have shown through LSA as well as DNS that locally stable regions can be introduced by suitably choosing viscosity profile and injection velocity to buoyantly unstable diffusive interface. Recently, Daniel and Riaz \cite{Daniel2014} used fixed interface and moving interface models to compare the theoretical predictions with the corresponding experimental observations \citep{Backhaus2011}. These authors showed, through an LSA, that in the absence of displacement the onset time increases when the dynamic viscosity increases with the depth. On the other hand, when the more viscous fluid displaces the less viscous one from above the onset time depends non-monotonically on the viscosity contrast between the two fluids. In both these studies LSA was performed under a quasi-steady state approximation, which has its own drawback, since it does not capture possible transient growth of the linearly unstable modes and hence fails to predict the onset of instability accurately. For flows with unsteady base-state, transient growth of the perturbations is possible in hydrodynamic stability problems driven solely by buoyancy force \citep{Rapaka2008} or viscous force \citep{Hota2015a}. 

Discussion of the above-mentioned literature reveals that transient growth plays important role on the onset of instability. Therefore, it is essential to discuss an LSA without quasi-steady state approximation, which not only captures the onset time more accurately, but also represents the physics appropriately. Recently, Hota {\it et al.} \cite{Hota2015b} have discussed an LSA based on an initial value problem (IVP) approach using a Fourier pseudo-spectral method. This IVP based LSA method captures the diffusion dominated region, which was never captured before using quasi-steady state approximation. In this context, we present an LSA \citep{Hota2015b} and DNS \citep{Tan1988} using a Fourier pseudo-spectral method to analyse the influence of viscosity contrast on buoyantly unstable miscible fluids in vertical porous media when the dynamic viscosity of the upper fluid is more or less than that of the lower fluid. From linear stability results it is identified that in the absence of displacement, viscosity contrast of either kind delays the onset of instability of gravitationally unstable miscible fluids. We further identify two different dynamical regimes in which the instability is dominated either by the viscous force or the buoyancy force. Our results pave the way to new insights into the influence of viscosity contrast on fingering instability in a buoyantly unstable miscible system. 

\begin{figure}
\centering
\includegraphics[width=4in, keepaspectratio=true, angle=0]{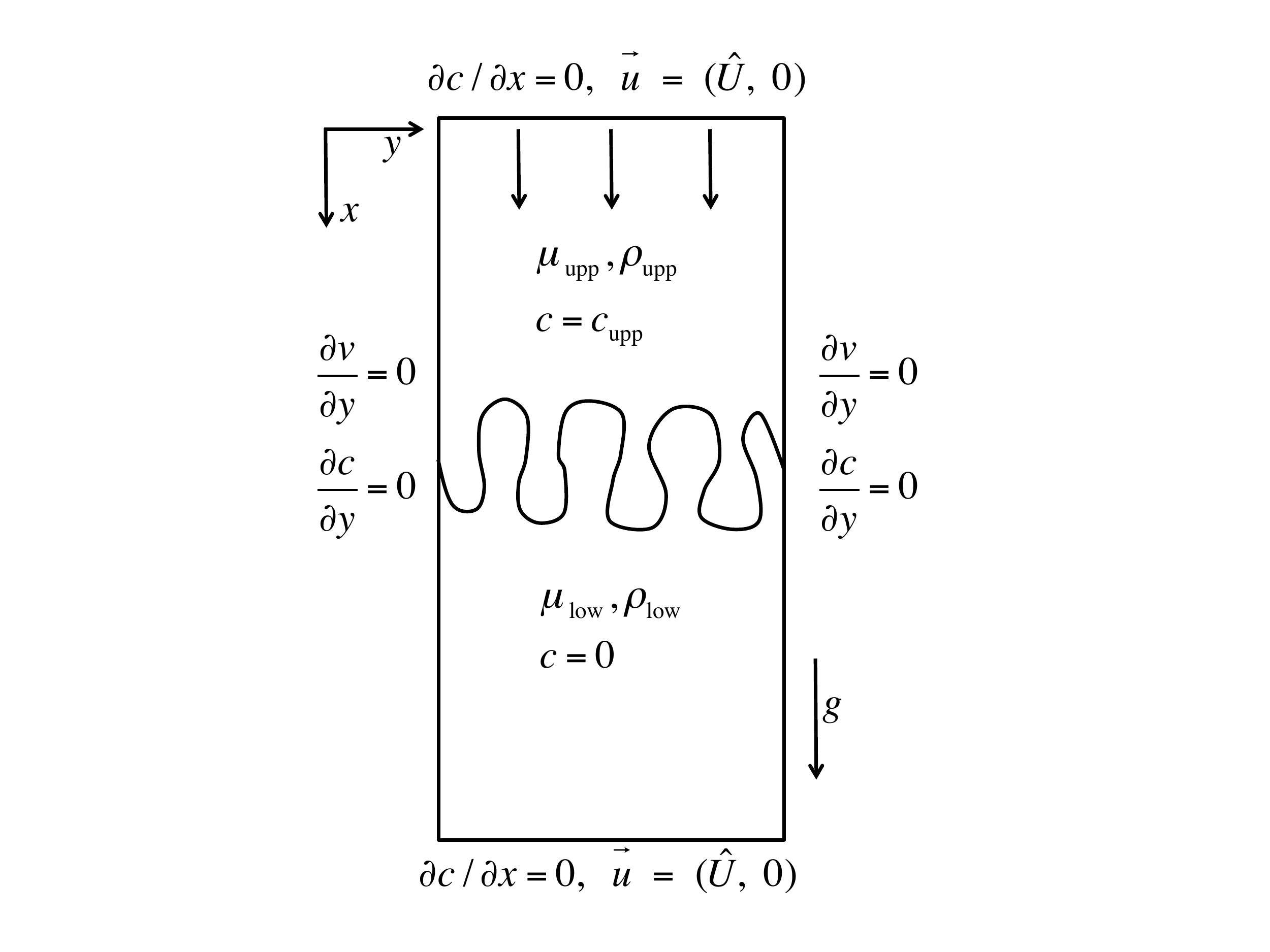}
\caption{Schematic of the flow. }\label{fig:schematic}
\end{figure}

\section{Problem formulation}\label{sec:PF}
\subsection{Governing equations}\label{subsec:GE}
Let us consider the displacement of a fluid of dynamic viscosity $\mu_{\rm low}$ and density $\rho_{\rm low}$ by another fluid of dynamic viscosity $\mu_{\rm upp}$ and density $\rho_{\rm upp}~ (> \rho_{\rm low})$ in a vertical porous medium, where the subscripts `low' and `upp' correspond to the lower and upper fluids, respectively. The fluids are assumed to be incompressible and miscible with each other. We further assume that the displacing fluid is injected at a uniform speed $\hat{U}$ vertically downward as shown in Fig. \ref{fig:schematic}. The dynamic viscosity of the fluid varies with a solute concentration, $c$, which satisfies a convection-diffusion equation. The fluid velocity can be determined in terms of the Darcy's law, which is a mathematical analogous to the flow equation in 2D homogeneous porous media. Hence, the governing equations can be written as, 
\begin{eqnarray}
\label{eq:cont}
& & \nabla\cdot\underline{u} = 0 \\
\label{eq:Darcy}
& & \nabla p = -\frac{\mu(c)}{\kappa}\underline{u} + \rho(c)\underline{g}, \\
\label{eq:CV}
& & \partial_t c + \underline{u}\cdot\nabla c = D \nabla^2 c,
\end{eqnarray}
where $\kappa$ is the permeability of the porous media, $\underline{g} = (g,0)$ with $g$ being the gravitational acceleration and $D$ is the isotropic dispersion co-efficient. 

\subsection{Dimensionless formulation}\label{subsec:DA}
In order to obtain dimensionless equations we use $L_{\rm ch} = V_{\rm ch}/D$ and $\tau_{\rm ch} = V_{\rm ch}/D^2$ as the characteristic length and time scales, respectively, where $V_{\rm ch} = |\Delta \rho|\kappa g/\mu_{\rm ch}$ is the buoyancy induced velocity, and $\mu_{\rm ch}$ is the characteristic viscosity. The characteristic pressure, concentration and density are taken to be $\mu_{\rm ch}D/\kappa, c_{\rm upp}$ and $|\Delta\rho| = |\rho_{\rm upp} - \rho_{\rm low}|$, respectively. The related dimensionless equations in a Lagrangian frame of reference moving with a dimensionless velocity, $U = \hat{U}/V_{\rm ch}$, are 
\begin{eqnarray}
\label{eq:cont_dim}
& & \nabla \cdot \underline{u} = 0 \\
\label{eq:Darcy_dim}
& & \nabla p = -\mu(c)\left(\underline{u} + Ue_x\right) + \rho(c)e_x, \\
\label{eq:CV_dim}
& & \partial_t c + \underline{u}\cdot\nabla c = \nabla^2 c, 
\end{eqnarray}
where $e_x$ is the unit vector in the $x$-direction. We assume that the density varies linearly with concentration, such that the dimensionless density profile is given as $\rho(c) = c$, 
\begin{equation}
\label{eq:viscosity}
\mu(c) = e^{Rf(c)}, 
\end{equation}
where $f(c)$ is a linear function of $c$ and the log-mobility ratio $R$ represents the natural logarithm of the ratio of the dynamic viscosities of two fluids. The explicit form of $f(c)$ and $R$ depend on the characteristic viscosity $\mu_{\rm ch}$. We define $\mu_{\rm ch}$, and the corresponding $f(c)$ and $R$ appropriately in \S \ref{sec:RD} while discussing the numerical results obtained from LSA and DNS. 

\subsection{Initial and boundary conditions}\label{subsec:IBC}
A description of appropriate initial and boundary conditions makes the mathematical formulation of the above problem complete. In the Lagrangian frame of reference the initial conditions for the concentration and velocity can be written as,
\begin{equation}
\label{eq:IC}
c = \left\{\begin{array}{lc}
1, & \mbox{for} ~~~ x < 0 \\
0, & \mbox{for} ~~~ x \geq 0
\end{array}
\right.,
~~~ \mbox{and} ~~~ 
\underline{u} = (0, 0).
\end{equation}
Along the longitudinal boundaries we have $\underline{u} = (0,0)$ and $\partial c/\partial x \to 0$ as $x \to \pm \infty$, while the transverse boundary conditions are $ \partial v/\partial y = 0$ (constant pressure cf. \cite{Nield}) and $\partial c/\partial y \to 0$. 

\section{Linear stability analysis}\label{sec:LSA}
In this section we discuss linear stability analysis of fingering instabilities driven by both viscosity and density contrasts in miscible displacements. 

\subsection{Stream function formulation}\label{subsec:SFF}
For a two dimensional flow the continuity equation can be satisfied identically by introducing a stream function, $\psi(x,y,t)$, such that $u = \partial \psi/\partial y$ and $v = -\partial \psi/\partial x$. Taking curl of equation (\ref{eq:Darcy_dim}) and representing velocities in terms of the stream function, equations (\ref{eq:Darcy_dim}) and (\ref{eq:CV_dim}) are recast as, 
\begin{eqnarray}
\label{eq:SF1}
& & \nabla^2\psi = -R\mathcal{D}f(c)\left(\nabla c \cdot \nabla \psi + U \partial_y c\right) + \frac{1}{\mu(c)}\partial_y c, \\
\label{eq:SF2}
& & \partial_t c + (\partial_y \psi) (\partial_x c) - (\partial_x \psi) (\partial_y c) = \nabla^2 c, 
\end{eqnarray} 
where $\mathcal{D} \equiv \mbox{d}/\mbox{d}c$. 

\subsection{Linearized perturbation equations}\label{subsec:LPE}
For base-state flow we assume $\underline{u}_b = 0$ that implies $\psi_b$ to be constant, which can be assumed to be equal to $0$ without any loss of generality. We also assume that the base-state concentration is homogeneous in the $y$-direction, i.e. $c_b = c_b(x,t)$. Under these assumptions base-state flow is given by decaying error function solution of step-like initial concentration profile, i.e., $c_b(x,t) = 0.5\;{\rm erfc}(x/2\sqrt{t})$. Introduce infinitesimal perturbations such that $c(x,y,t) = c_b(x,t) + c'(x,y,t)$ and $\psi(x,y,t) = \psi_b + \psi'(x,y,t)$, etc., and substitute these in equations (\ref{eq:SF1}) and (\ref{eq:SF2}) to obtain,
\begin{eqnarray}
\label{eq:SF_total1}
& & \partial_t c_b + \partial_t c' + \left( \partial_x c_b + \partial_x c' \right) \partial_y \psi' - (\partial_x \psi') (\partial_y c') = \nabla^2 c_b + \nabla^2 c', \\
\label{eq:SF_total2}
& & \nabla^2\psi' = -R\mathcal{D}f(c_b + c')\left[\left(\nabla c' + (\partial_x c_b) e_y\right)\cdot \nabla \psi' + U \partial_y c'\right] + \frac{1}{\mu(c_b + c')} \partial_y c', 
\end{eqnarray}
with $e_y$ being the unit vector in the $y$-direction. Subtracting the base-state equations from equations (\ref{eq:SF_total1}) and (\ref{eq:SF_total2}) we obtain the following coupled nonlinear partial differential equations in terms of the perturbation quantities $c'$ and $\psi'$ as,
\begin{eqnarray}
\label{eq:SF_perturb1}
& & \partial_t c' + \left(\partial_x c_b + \partial_x c'\right)\partial_y \psi' - (\partial_y c') (\partial_x \psi') = \nabla^2 c', \\
\label{eq:SF_perturb2}
& & \nabla^2\psi' = -R\mathcal{D}f(c)\left[\left(\nabla c' + (\partial_x c_b) e_y\right)\cdot \nabla \psi' + U \partial_y c'\right] + \frac{1}{\mu(c_b + c')} \partial_y c'. 
\end{eqnarray}
Linearizing these equations in terms of the perturbation quantities we obtain,
\begin{eqnarray}
\label{eq:SF_linear1}
& & \partial_t c' + \partial_x c_b \partial_y \psi' =\nabla^2c', \\
\label{eq:SF_linear2}
& & \nabla^2\psi' = -R\mathcal{D}f(c)\left[(\partial_x c_b) (\partial_x \psi') + U \partial_y c'\right] + \frac{1}{\mu(c_b)}\partial_y c'. 
\end{eqnarray}
We solve equations (\ref{eq:SF_linear1}) and (\ref{eq:SF_linear2}) using a pseudo-spectral method to obtain the spatio-temporal evolution of the perturbation quantities, $c'(x,y,t)$ and $\psi'(x,y,t)$ and calculate the growth rates associated with concentration and velocity perturbations \citep{Kumar1999, Hota2015b}, 
\begin{eqnarray}
\label{eq:GR}
\sigma_{c} = \frac{1}{2E_{c'}}\frac{\mbox{d}E_{c'}}{\mbox{d}t}, ~~~ \sigma_{V} = \frac{1}{2E_{v'}}\frac{\mbox{d}E_{v'}}{\mbox{d}t}, ~~~ \sigma = \frac{1}{2E}\frac{\mbox{d}E}{\mbox{d}t}, 
\end{eqnarray}
from the amplification measures defined as \citep{Hota2015b},
\[E_{c'} = \int\int \left(c'\right)^2\mbox{d}x\mbox{d}y, ~~~ E_{v'} = \int\int \left[\left(\partial_y \psi' \right)^2 + \left(\partial_x \psi'\right)^2\right]\mbox{d}x\mbox{d}y, ~~~ E = E_{c'} + E_{v'}. \]
Following Hota {\it et al.} \cite{Hota2015b} we have used $\sigma$ (defined in equation (\ref{eq:GR})) to quantify the growth rate of disturbances and the onset of instability. 

\section{Results and discussion}\label{sec:RD}
In this section we discuss the numerical results and their physical interpretations for different flow parameters. In the absence of viscosity contrast convective instability is featured at the miscible interface when a heavier fluid is placed above a lighter fluid. This hydrodynamic instability is broadly known as density fingering (DF) in the literature. How is this convective instability modified with the viscosity contrast between the underlying fluid? Here we investigate the influence of the viscosity contrast on buoyantly unstable miscible fluids for both $U = 0$ and $U \neq 0$, respectively, in \S \ref{subsec:U_0} and \S \ref{subsec:U_not_0}. 

\subsection{Effect of viscosisty contrast in the absence of displacement}\label{subsec:U_0}

Daniel and Riaz \cite{Daniel2014} presented fixed interface and moving interface methods to compare the natural convections ($U = 0$) when the dynamic viscosity of the upper fluid is more or less than that of the lower fluid. With the help of moving interface method they showed that the onset of instability is delayed when the dynamic viscosity increases with depth compared to the case of viscosity matched fluids. On the other hand, the instability sets in earlier when the dynamic viscosity decreases with depth. This is in contrary to the situation of the classical viscous fingering instability in neutrally buoyant fluids. 

\begin{figure}
\centering
(a) \hspace{2.5in} (b) \\
\includegraphics[width=2.7in, keepaspectratio=true, angle=0]{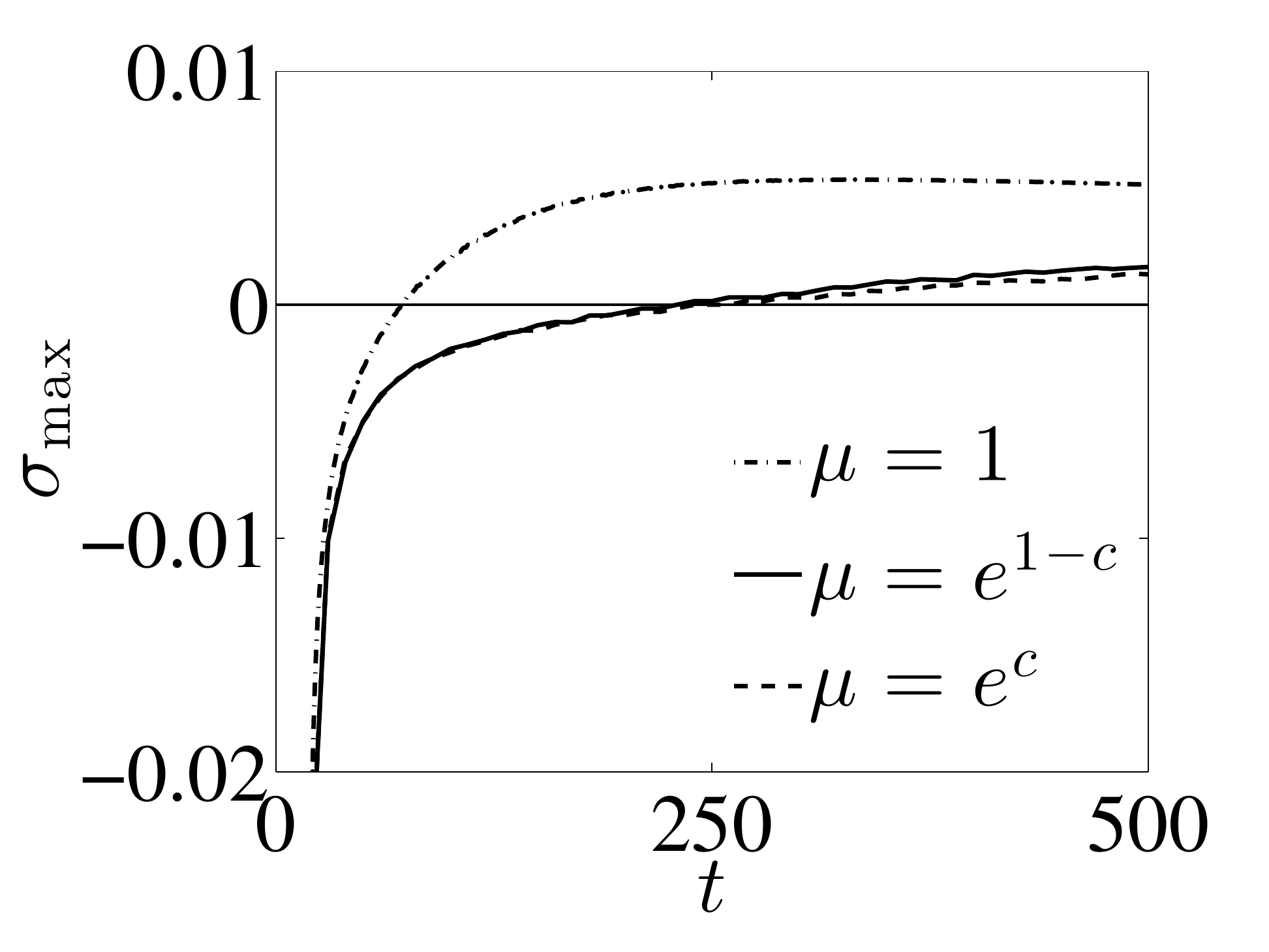}
\includegraphics[width=2.7in, keepaspectratio=true, angle=0]{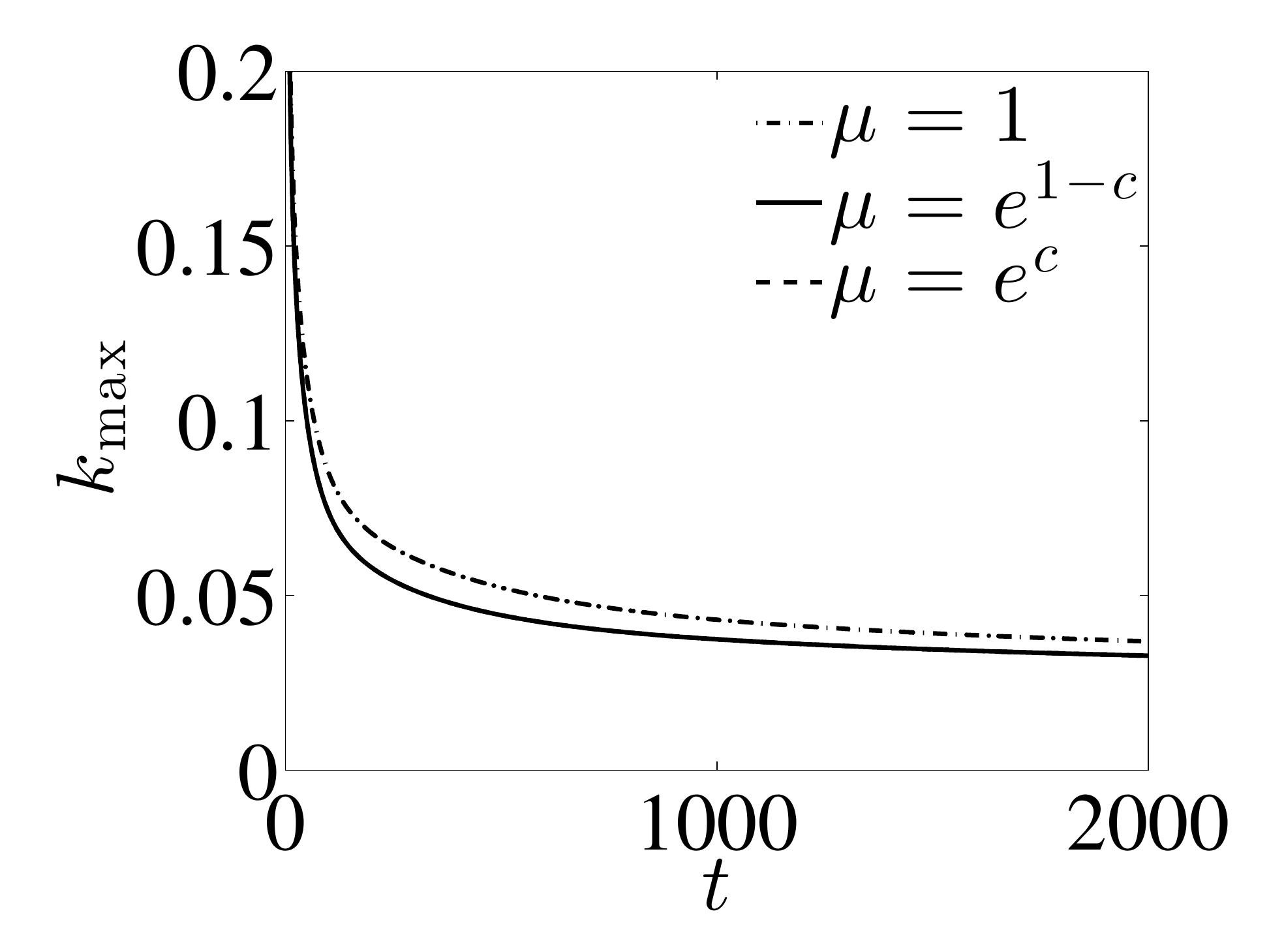}
\caption{(a) Maximum growth rate, $\sigma_{\rm max}$, and (b) dominant wavenumber, $k_{\rm max}$ for $U = 0$. }\label{fig:LSA_U_0}
\end{figure}

We revisit the problem of moving interface (cf. \cite{Daniel2014}) by choosing the characteristic viscosity $\mu_{\rm ch} = \mu_l$, where $\mu_l$ corresponds to the viscosity of the less viscous fluid. For example, $\mu_{\rm ch} = \mu_{\rm upp}$ (or $\mu_{\rm ch} = \mu_{\rm low}$) when $\mu_{\rm upp} < \mu_{\rm low}$ (or $\mu_{\rm low} < \mu_{\rm upp}$). In this case the linear function $f(c)$ and the log-mobility ratio $R$ are defined as,
\begin{equation}
\label{eq:f}
f(c) = \left\{
\begin{array}{lc}
1-c, & \mbox{if} ~~~ \mu_{\rm upp} < \mu_{\rm low} \\
c, & \mbox{if} ~~~ \mu_{\rm upp} > \mu_{\rm low} 
\end{array}
\right., ~~~ R = \ln\left(\frac{\mu_m}{\mu_l}\right),
\end{equation}
respectively, where $\mu_m$ corresponds to the viscosity of the more viscous fluid. Linear stability analysis is performed for $R = 1$ and $U = 0$. The temporal evolution of the maximum growth rate, $\sigma_{\rm max}$, and the dominant wave number, $k_{\rm max}$, of the perturbation quantities are shown in figure \ref{fig:LSA_U_0}. Figure \ref{fig:LSA_U_0}(a) depicts that the onset happens at the earliest for the classical DF case compared to the situations when the viscosity of the two fluids are different. Moreover, the temporal evolution of the growth rates obtained for different viscosity contrasts are visually indistinguishable. The physical explanation for different onset time of instability in flow with or without viscosity contrast in buoyantly unstable miscible fluids can be presented in terms of the instantaneous vorticity perturbation field as discussed by Daniel and Riaz \cite{Daniel2014}. From figure \ref{fig:LSA_U_0}(b) it is identified that for all time the most unstable wave numbers are larger in the absence of the viscosity contrast as compared to the situation when the viscosity of the two fluids are different. Thus we conclude that for $U = 0$ viscous force induces stability to the classical DF instability in a vertical porous medium. 

From \S \ref{subsec:DA} it is clearly observed that all the characteristic scales are derived using the characteristic viscosity. Daniel and Riaz \cite{Daniel2014} chose $\mu_{\rm ch} = \mu_{\rm upp}$ for both the more and less viscous upper fluid, so that the viscosity-concentration relation (\ref{eq:viscosity}) takes the form $\mu(c) = e^{R(1-c)}$, where $R = \ln(\mu_{\rm low}/\mu_{\rm upp})$. Therefore, $R > 0$ (or $R < 0$) corresponds to the less (or more) viscous upper fluid. Such a choice of the characteristic viscosity generates different length, time and velocity scales for the respective problem of more or less viscous upper fluid. Writing the characteristic length, time, velocity and dynamic viscosity, corresponding to $R > 0$ as $L_{\rm ch}^+, \tau_{\rm ch}^+, V_{\rm ch}^+$ and $\mu_{\rm ch}^+$, and those for $R < 0$ as $L_{\rm ch}^-, \tau_{\rm ch}^-, V_{\rm ch}^+-$ and $\mu_{\rm ch}^-$, respectively, we have 
\begin{equation}
\label{eq:relation}
\mu_{\rm ch}^-/\mu_{\rm ch}^+ = 1/\alpha, ~~~ V_{\rm ch}^-/V_{\rm ch}^+ = \alpha,~~~ L_{\rm ch}^-/L_{\rm ch}^+ = 1/\alpha, ~~~ \tau_{\rm ch}^-/\tau_{\rm ch}^+ = 1/\alpha^2.
\end{equation} 

\begin{table}
\centering
\begin{tabular}{ccccccccccccc}
~ &\multicolumn{6}{c}{(i)} ~~~~ &\multicolumn{6}{c}{(ii)} \vspace{0.1cm} \\ 
$\alpha$ ~~~~ & $\mu_{\rm ch}$ & $R$ & $\mu(c)$ & Vel. & Len. & Time ~~~~  & $\mu_{\rm ch}$   & $R$ & $\mu(c)$ & Vel. & Len. & Time \\ \\
$10 ~ (> 1)$ ~~~~ & $\mu_{\rm upp}$  & 2.3 & $e^{R(1-c)}$ & $U$ & $L$ & $t$ ~~~~ & $\mu_{\rm upp}$ & 2.3 & $e^{R(1-c)}$ & $U$ & $L$ & $t$ \\
$ 0.1 ~ (< 1)$ ~~~~ &$\mu_{\rm low}$ & 2.3 & $e^{Rc}$ & $U$ & $L$ & $t$ ~~~~ & $\mu_{\rm upp}$  & -2.3 & $e^{R(1-c)}$ & $U/\alpha$ & $\alpha L$ & $\alpha^2 t$  
\end{tabular}
\caption{The log-mobility ratio ($R$), dimensionless velocity (Vel.), dimensionless length (Len.) and dimensionless time (Time) corresponding to two different viscosity scales are shown for a given set of dimensional values. Here $U = \hat{U}/V^+_{\rm ch}, L = \hat{L}/L^+_{ch}$ and $t = \hat{t}/\tau^+_{\rm ch}$, with $\hat{\cdot}$ being the dimensional value of the respective variables. }
\label{table:visco1}
\end{table}

Thus comparison of the onset of instability and fingering dynamics between these two cases should be performed by suitably choosing the characteristic viscosity. In the present analysis $\mu_l$ is chosen as the characteristic viscosity irrespective of whether the upper fluid is more viscous or less viscous than the lower fluid. Using such a viscosity scaling the length, time and velocity scales of the two fluid flow problems corresponding to a more or less viscous fluid at the top remain the same. For a given set of dimensional values of the displacement velocity, domain length and time, respective dimensionless values corresponding to $\mu_{\rm ch} = \mu_{\rm upp}$ \citep{Manickam1995, Daniel2014} and $\mu_{\rm ch} = \mu_l$ are presented in table \ref{table:visco1}. It is identified that corresponding to $\mu_{\rm ch} = \mu_{\rm upp}$, the dimensionless values obtained for a more viscous fluid at the top are different from those when the less viscous fluid is at the top. A simple rescaling of the dimensionless displacement velocity, length and time of the problem is essential to compare between the cases of more ($R < 0$) or less ($R > 0$) viscous fluid at the top as shown in table \ref{table:visco1}. However, Daniel and Riaz \cite{Daniel2014} used $U, L$ and $t$ as the dimensionless values for both $R > 0$ and $R < 0$ (see figures 4, 5 and 12a, etc. of \cite{Daniel2014}). This rescaling can be avoided with $\mu_{\rm ch} = \mu_l$.

\begin{figure}
\centering
\includegraphics[width=2.5in, keepaspectratio=true, angle=0]{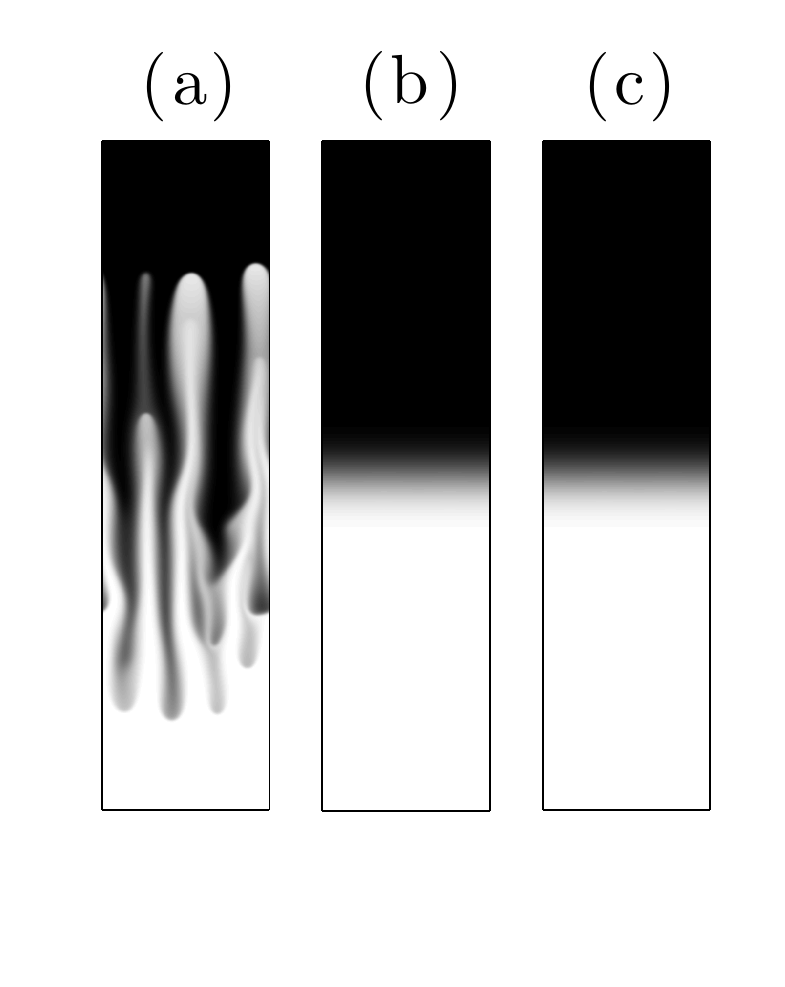}
\vspace{-0.5in}
\caption{Spatial distribution of the solute concentration for $U = 0$: (a) $\mu(c) = e^{-2.3(1-c)}$ at $t = 2 \times 10^3$, (b) $\mu(c) = e^{2.3c}$ at $t = 10^4$ and (c) $\mu(c) = e^{-2.3(1-c)}$ at $\alpha^2t = 10^4/10^2 = 10^2$. }\label{fig:DNS1}
\end{figure}

In order to understand the influence of viscosity scaling we consider the problem of viscous stabilization of buoyantly unstable miscible layers, i.e. $\alpha = \mu_{\rm low}/\mu_{\rm upp} < 1$, such that $\mu_{\rm ch} = \mu_{\rm upp}$ results $\mu(c) = e^{R(1-c)}$ with $R = \ln(\mu_{\rm low}/\mu_{\rm upp}) <0 $, while $\mu_{\rm ch} = \mu_{\rm low}$ corresponds to $\mu(c) = e^{Rc}$ with $R = \ln(\mu_{\rm upp}/\mu_{\rm low}) > 0$. Therefore, suitably choosing the length, time and velocity of the flow problems, one would expect to identify the same results for $R < 0$ and $R > 0$. In order to illustrate this fact, we choose $U = 0$ and $\alpha = 0.1$ so that $|R| \approx 2.3$ and perform DNS using a Fourier pseudo-spectral method \citep{Tan1988} to support our theoretical analysis. The obtained numerical results are depicted in figures \ref{fig:DNS1}(a-c).  Figure \ref{fig:DNS1}(a), which corresponds to the viscosity scaling applied by Daniel and Riaz \cite{Daniel2014}, i.e. $\mu(c) = e^{-2.3(1-c)}$, shows that the diffusive interface features fingers. On the other hand, DNS results corresponding to $\mu_{\rm ch} = \mu_{\rm low}$ (i.e. $\mu(c) = e^{2.3c}$) depict pure diffusive expansion of the miscible interface (see figure \ref{fig:DNS1}(b)). We also perform DNS corresponding to $\mu(c) = e^{-2.3(1-c)}$ with rescaled length and time scales according to the above-mentioned relations (see equation \ref{eq:relation}). Spatial distribution of the solute concentration at $\alpha^2t = 10^2$ is shown in figure \ref{fig:DNS1}(c), which is identical to figure \ref{fig:DNS1}(b). From linear theory we identified that the dynamics of the systems for more and less viscous upper fluid are indistinguishable. The analogous results can also be shown in the nonlinear regime through DNS. Thus we conclude that the comparative study presented by Daniel and Riaz \cite{Daniel2014} is inappropriate, since the length, time and velocity scales associated with the problems related to more or less viscous upper fluid are different in their study. 

\begin{figure}
\centering
(a) \hspace{2.5in} (b) \\
\includegraphics[width=2.7in, keepaspectratio=true, angle=0]{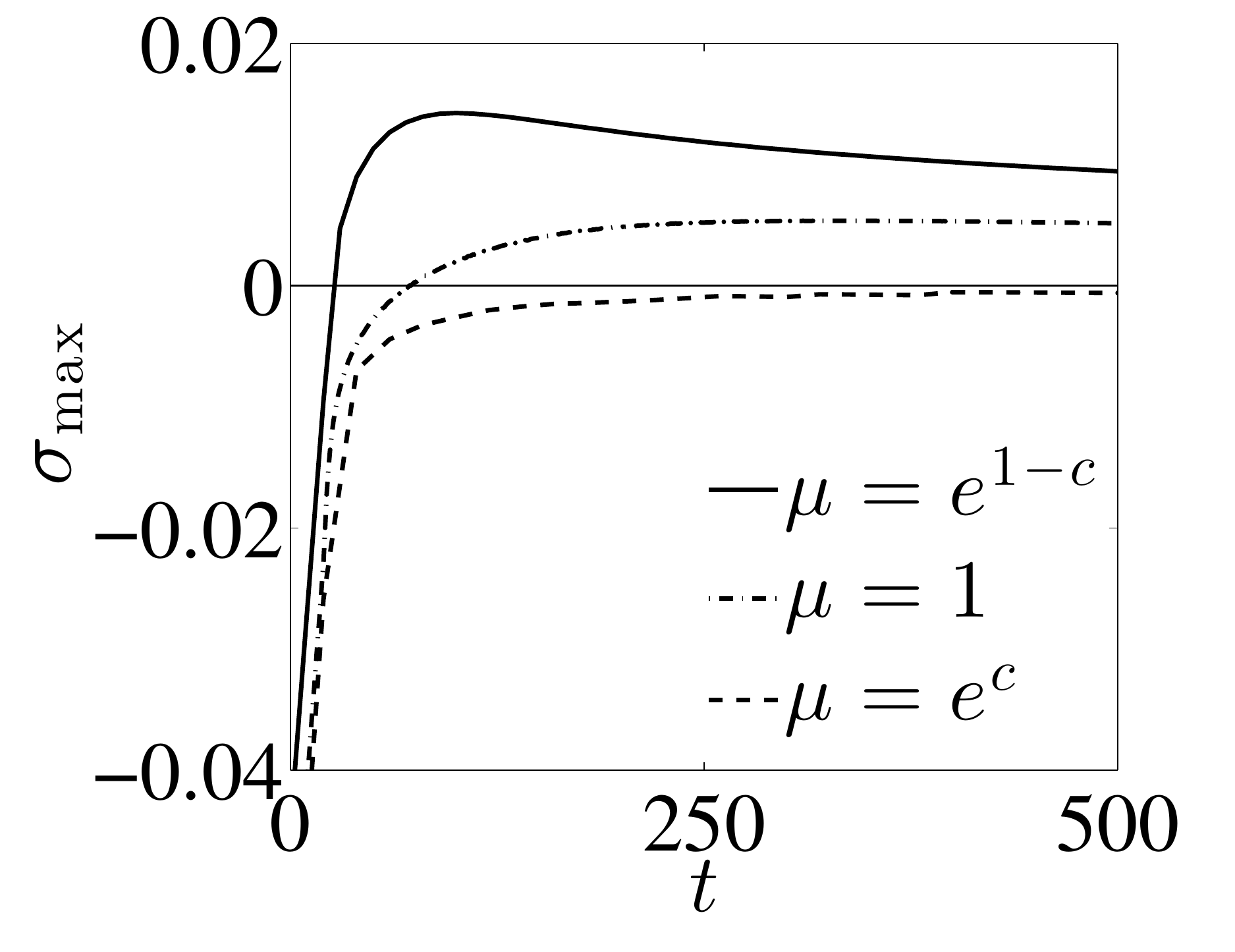}
\includegraphics[width=2.7in, keepaspectratio=true, angle=0]{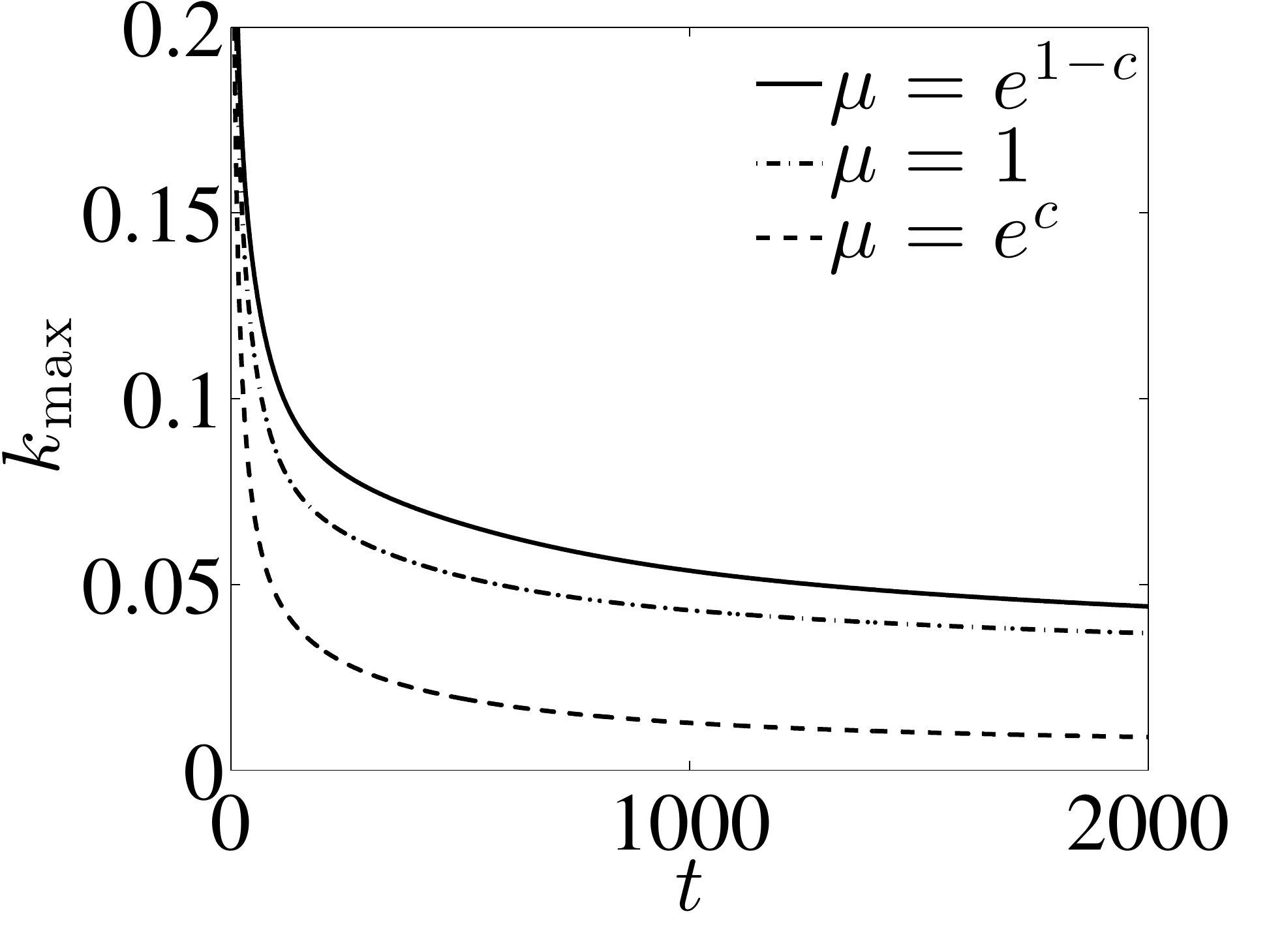}
\caption{(a) Maximum growth rate, $\sigma_{\rm max}$, and (b) dominant wavenumber, $k_{\rm max}$ for $U = 1$. }\label{fig:LSA_U_1}
\end{figure}

\subsection{Influence of the displacement velocity}\label{subsec:U_not_0}
Next, we consider the displacement of the lower fluid by the upper one. The influence of both the stable and unstable viscosity contrasts are discussed through LSA as well as DNS. Figure \ref{fig:LSA_U_1} shows the temporal evolution of $\sigma_{\rm max}$ and $k_{\rm max}$ for the same parameters as those of figure \ref{fig:LSA_U_0}, except for $U = 1$. It is identified that the dynamics of buoyancy induced instability in viscosity matched fluids remains unaffected with the dimensionless displacement velocity, hence the onset of instability are the same for $U = 0$ (dash-dotted line in figure \ref{fig:LSA_U_0}(a)) and $U = 1$ (dash-dotted line in figure \ref{fig:LSA_U_1}(a)) when $\mu(c) = 1$. Figure \ref{fig:LSA_U_1}(a) depicts that instability sets in earlier when the less viscous heavier fluid at the top displaces the more viscous lighter fluid at the bottom. The unfavourable viscosity contrast coupled with buoyancy force enhances the instability, which is readily evident from the fact that for all time the growth rate of the perturbations for $\mu(c) = e^{1-c}$ are larger than the respective values when $\mu(c) = 1$ (see figure \ref{fig:LSA_U_1}(a)). On the other hand, for $\mu(c) = e^{c}$ the favourable viscosity contrast acts against the instability induced by the buoyancy force and the displacement becomes stable (see figure \ref{fig:LSA_U_1}(a)). Figure \ref{fig:LSA_U_1}(b) illustrates that at a given time and for the parameter values scanned here, the most unstable wave number $k_{\rm max}$ is the largest for $\mu(c) = e^{1-c}$ and smallest for $\mu(c) = e^c$. In other words, at a given time $k_{\rm max}$ increases with $\alpha$. Such influences of the viscosity contrast on the stability of buoyantly unstable miscible fluids are similar to those in neutrally buoyant fluids, i.e. in the case of VF instability \citep{Homsy1987}. In summary, for neutrally buoyant as well as buoyantly unstable miscible fluids an unfavourable viscosity contrast enhances the instability, whereas a favourable viscosity contrast weakens the instability, when the upper fluid displaces the lower one. 

\begin{figure}
\centering
\includegraphics[width=5in, keepaspectratio=true, angle=0]{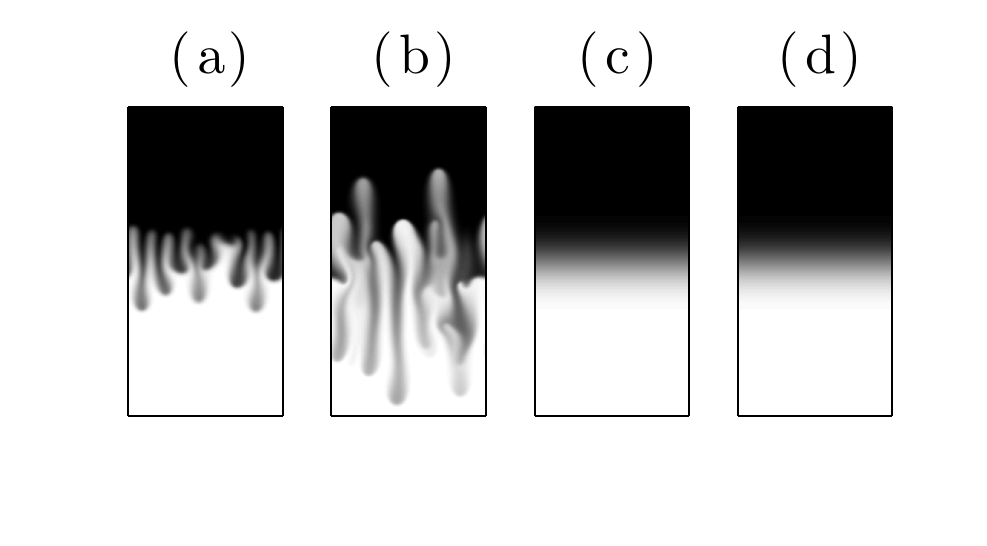}
\vspace{-0.2in}
\caption{Spatial distribution of the solute concentration: (a) $\mu(c) = e^{2.3(1-c)}, U = 0.5$ at $t = 2 \times 10^3$, (b) $\mu(c) = e^{-2.3(1-c)}, U = 0.5$ at $t = 2 \times 10^3$, (c)  $\mu(c) = e^{2.3c}, U = 0.5$ at $t = 10^4$ and (d) $\mu(c) = e^{-2.3(1-c)}, \alpha U = 0.5 \times 10 = 5$ at $\alpha^2t = 10^4/10^2 = 100$. }\label{fig:DNS2}
\end{figure}

In \S \ref{subsec:U_0} we show, for $U = 0$, the dimensionless length and time should be chosen wisely to compare between $R > 0$ and $R < 0$. Here we continue similar analysis for $U \neq 0$. As an example we choose $U = 0.5$, and perform DNS when the dynamic viscosity of the upper fluid is more or less than the lower one. Following Daniel and Riaz \cite{Daniel2014} we choose $\mu_{\rm ch} = \mu_{\rm upp}$ and compare the dynamics of less viscous upper fluid, i.e. $\mu(c) = e^{2.3(1-c)}$, (see figure \ref{fig:DNS2}(a)) with that of the more viscous upper fluid, i.e. $\mu(c) = e^{-2.3(1-c)}$ (see figure \ref{fig:DNS2}(b)). Counter-intuitive results, that the displacement of a less viscous fluid by a more viscous one features stronger instability than the displacement of a more viscous fluid by a less viscous one, are identified. Next we take $\mu_{\rm ch} = \mu_l$, such that the displacement of more viscous fluid at the bottom by less viscous fluid at the top is represented by $\mu(c) = e^{2.3c}$ and the corresponding spatial distribution of the solute concentration is presented in figure \ref{fig:DNS2}(c). It depicts that the miscible interface features only diffusive expansion, which was also mentioned by Manickam and Homsy \cite{Manickam1995} through LSA. Further, we  perform numerical simulations for $\mu(c) = e^{-2.3(1-c)}$ with rescaled dimensionless velocity, length and time as mentioned in table \ref{table:visco1}. The result obtained from DNS is depicted in figure \ref{fig:DNS2}(d), which is indistinguishable from figure \ref{fig:DNS2}(c). This signifies the essence of an appropriate viscosity scaling while comparing the influence of more or less viscous fluid at the top on the dynamics of a buoyantly unstable miscible interface. 

Discussion of figures \ref{fig:LSA_U_0}-\ref{fig:DNS2} illustrates that a simpler and convenient choice of $\mu_{\rm ch}$ is the dynamic viscosity of the less viscous fluid, i.e. $\mu_l$, which preserves the same dimensionless length, time, velocity, etc. for more or less viscous upper fluid. In the rest of the paper we choose $\mu_{\rm ch} = \mu_l$, such that the viscosity-concentration relation is given by (\ref{eq:f}). 

\begin{figure}
\centering
\includegraphics[width=3.5in, keepaspectratio=true, angle=0]{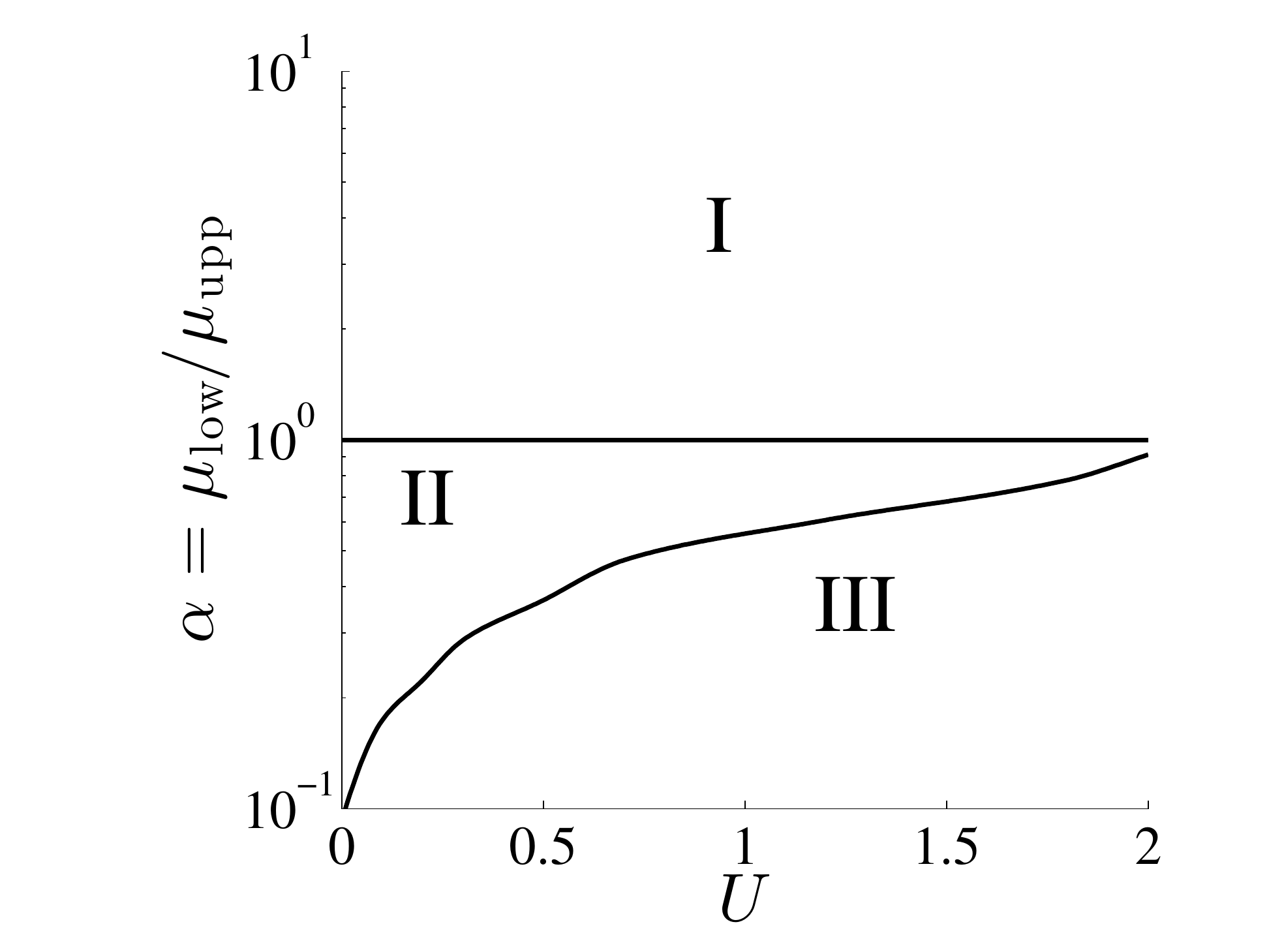}
\caption{Different stability regions in phase space spanned by the displacement velocity, $U$, and viscosity ratio, $\alpha = \mu_{\rm low}/\mu_{\rm upp}$. Regions I and II correspond to the instabilities dominated by buoyancy and viscosity, respectively, while the stable region is denoted by III. }\label{fig:phase}
\end{figure}

The stability scenarios in the phase space spanned by the displacement velocity, $U$, and the mobility ratio, $\alpha = \mu_{\rm low}/\mu_{\rm upp}$, are shown in figure \ref{fig:phase}. The parameter space can be divided into three distinct regions depending on the stability characteristics. The buoyancy and viscosity dominated instability regions are denoted by I and II, respectively, and the stable region is represented by region III. For the viscosity matched fluids (i.e. $\alpha = 1$) the diffusive interface is buoyantly unstable and the growth rate of the perturbations are indistinct for all possible values of the dimensionless displacement velocity, $U$. For $\alpha > 1$, i.e. when a less viscous heavier fluid at the top displaces a more viscous lighter one at the bottom, the instability at the diffusive interface is driven by both the viscosity and density contrasts, and the instability increases with $\alpha$ as well as $U$. On the other hand, for $\alpha < 1$, i.e. for viscous stabilisation of buoyantly unstable diffusive interface, the instability becomes weak as $\alpha$ decreases or $U$ increases, and finally becomes stable when $U$ (or $\alpha$) is larger (or smaller) than a certain critical value. 

\begin{figure}
\centering
\includegraphics[width=3.5in, keepaspectratio=true, angle=0]{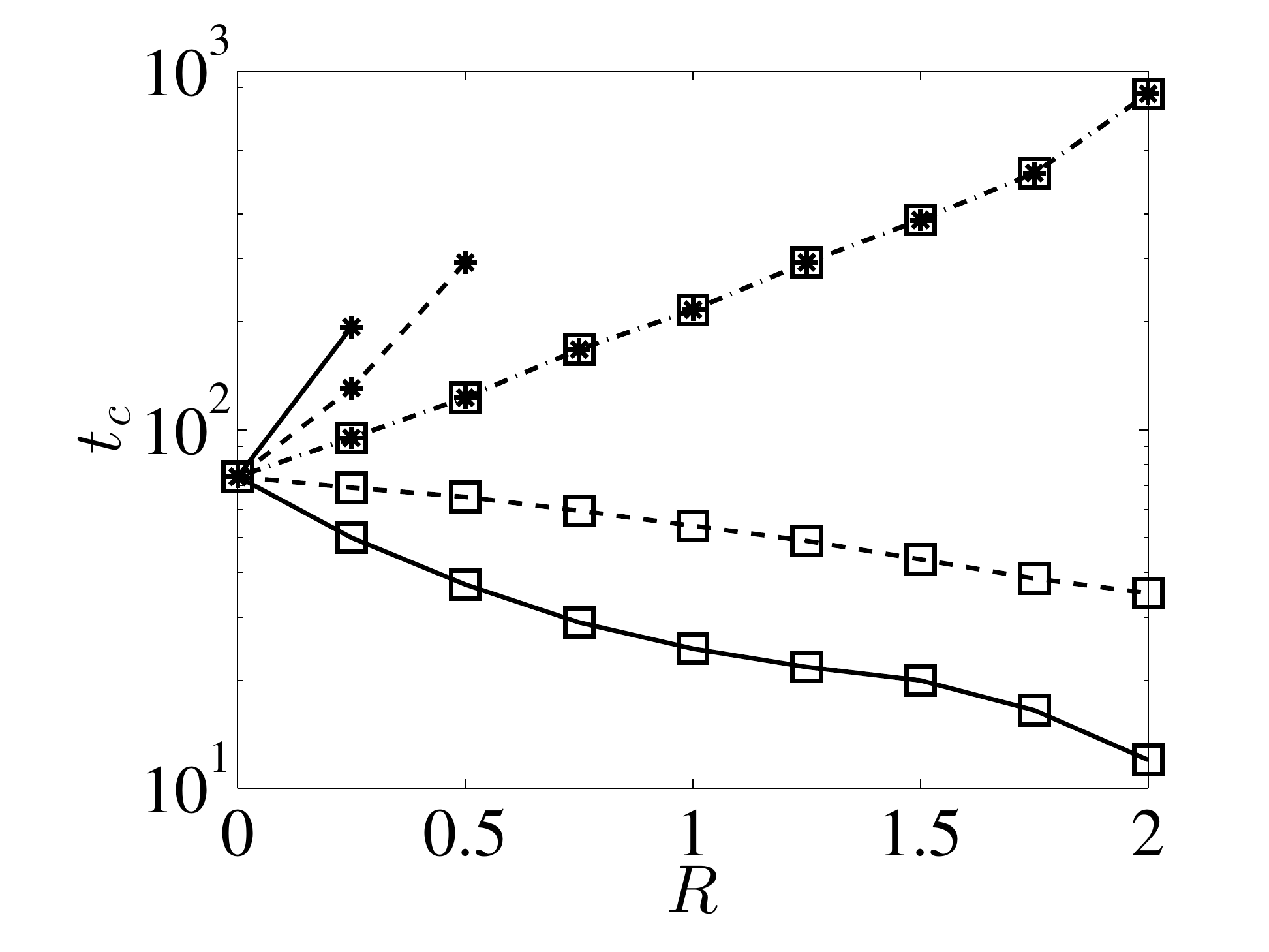}
\caption{Effect of viscosity contrast on the onset time, $t_c$, for different dimensionless displacement velocity, $U = 1$ (solid line), $U = 0.5$ (dashed line) and $U = 0$ (dash-dotted line). Square ($\square$) corresponds to $\mu(c) = e^{R(1-c)}$ and asterisk ($\ast$) corresponds to $\mu(c) = e^{Rc}$. }\label{fig:t_c}
\end{figure}

\begin{figure}
\centering
\includegraphics[width=3.5in, keepaspectratio=true, angle=0]{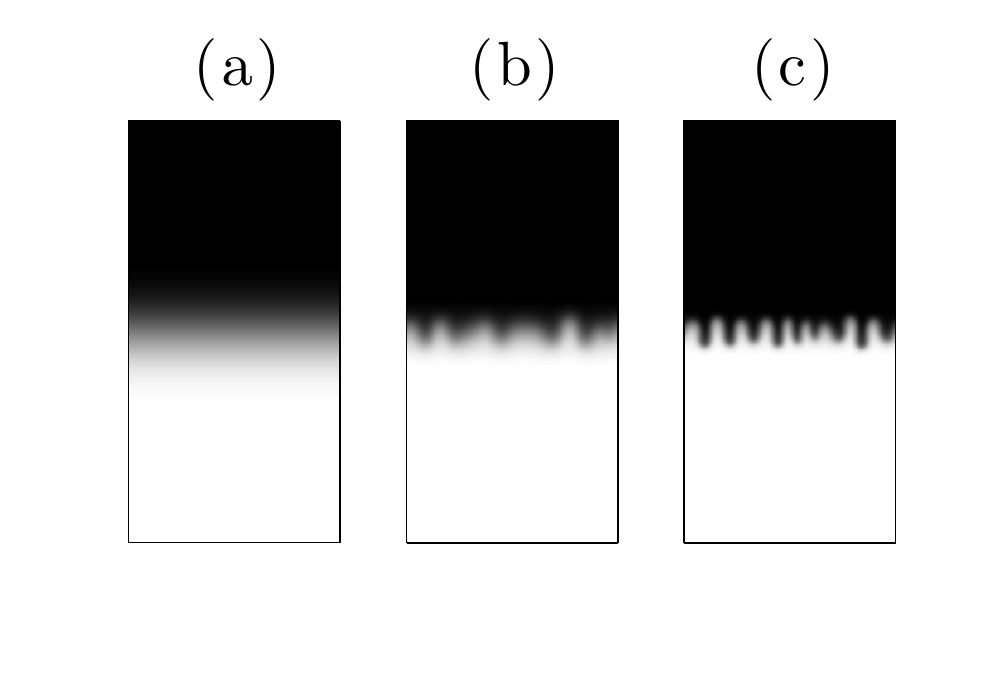}
\vspace{-0.5in}
\caption{Spatial distribution of the concentration, $c$, for $U = 1$: (a) $\mu(c) = e^{c}$ at $t = 10000$, (b) $\mu(c) = 1$ at $t = 2400$ and (c) $\mu(c) = e^{(1-c)}$ at $t = 1000$. }\label{fig:DNS3}
\end{figure}

The influence of the viscosity contrast on the onset of instability is depicted in figure \ref{fig:t_c}. This figure shows, in the absence of displacement ($U = 0$), the onset of instability delays with the viscosity contrast, irrespectively whether the heavier fluid is less viscous or more viscous. The present results are consistent with Daniel and Riaz \cite{Daniel2014} when a more viscous heavier fluid is placed over a less viscous lighter fluid. These authors used $\mu_{\rm upp}$ as the characteristic viscosity and showed, for $U = 0$ the onset time increases with mobility ratio $\alpha = \mu_{\rm low}/\mu_{\rm upp}$. More surprisingly, they found an early onset of instability with stable viscosity contrasts ($\alpha < 1$) and $U > 0$, compared to buoyantly unstable miscible interface between viscosity matched fluids, i.e. $\alpha = 1$ (see figure 12a of Daniel and Riaz \cite{Daniel2014}). This result is contradictory to the result of Manickam and Homsy \cite{Manickam1995}, who showed that a buoyantly unstable miscible interface can be stabilised by suitably choosing the viscosity contrast and the displacement velocity. The present LSA results successfully captures this phenomenon (see figure \ref{fig:LSA_U_1}). The onset of instability as a function of the log-mobility ratio, both for more and less viscous upper fluid, is depicted in figure \ref{fig:t_c}. It is identified that the instability sets in earlier with increasing $R$ when a less viscous fluid displaces a more viscous fluid. On the other hand, during the displacement of less viscous fluid by a more viscous one, onset time increase with $R$ and the displacement becomes completely stable after a threshold value of $R$, which depends on the injection velocity $U$. In order to confirm the present linear stability results, DNS are performed for the parameter values used in figure \ref{fig:LSA_U_1}. The spatial distribution of the concentration for displacements with and without viscosity contrast are shown in figure \ref{fig:DNS3}. This figure depicts that a buoyantly unstable diffusive interface of viscosity matched fluids (middle panel in figure \ref{fig:DNS3}) becomes stable when the dynamic viscosity of the upper fluid is more (left panel of figure \ref{fig:DNS3}). On the other hand, the instability becomes stronger when a less viscous fluid at the top displaces a more viscous fluid at the bottom (right panel of figure \ref{fig:DNS3}). 

\subsection{Non-orthogonality of eigenmodes and transient growth}\label{subsec:TG}
Here, we briefly discuss about the possible transient growth of the disturbances to the unsteady base state $c_b(x,t) = 0.5\;{\rm erfc}(x/2\sqrt{t})$. Manickam and Homsy \cite{Manickam1995} presented an LSA based on the modal analysis under the assumption of quasi-steady state approximation (QSSA). However, the QSSA modes are non-orthogonal \citep{Rapaka2008, Hota2015a} and thus these eigenvalues do not reveal the exotic transient behavior (Schmid2007). Further, Trefethen {\it et al.} \cite{Trefethen1993} reported that the transient growth in a stable linearised system has implications for the behaviour of the associated nonlinear system.

To quantify the degree of non-orthogonality of the eigenmodes obtained from modal analysis we study the linear stability of the unsteady base state $c_b(x,t)$ with respect to small wavelike perturbations of the form
\begin{equation}
\left( c', u'\right)(x,y,t) = \left( \phi_c, \phi _u\right)(x,t)\mbox{exp}({\rm i}ky),
\end{equation}
where ${\rm i} = \sqrt{-1}$, $k$ is the non-dimensional wave number in the $y$ direction, and $\phi_c(x,t), \phi_u(x,t)$ are time dependent  concentration and velocity perturbations, respectively. Following the standard procedure \citep{Homsy1987,Pramanik2015}, the linear stability equations in a similarity transformation $(\xi, t)$-domain can be written as an initial value problem (IVP) for $\phi_c, \phi_u$
\begin{eqnarray}
\label{eq:LSA1}
& & \frac{\partial \phi_c}{\partial t} = \left\{\left(\frac{1}{t}\frac{\partial^2}{\partial\xi^2} - k^2\right) + \frac{\xi}{2t}\frac{\partial}{\partial \xi} \right\}\phi_c - \frac{1}{\sqrt{t}}\frac{\text{d}c_b}{\text{d}\xi}\phi_u, \\
\label{eq:LSA2}
& & \left(\frac{\partial^2}{\partial\xi^2} + Rf'(c_b) \frac{\partial}{\partial\xi} - k^2t\right)\phi_u = k^2t\left(URf'(c_b) - \frac{\rho'(c_b)}{\mu_0}\right)\phi_c,
\end{eqnarray}
where $\xi = x/\sqrt{t}$ is the similarity variable. Finite difference approximation of the linearized operators followed by elimination of $\phi_u$ from equations \eqref{eq:LSA1} and \eqref{eq:LSA2} yields a nonautonomous system of ordinary differential equations, 
\begin{equation}
\label{eq:IVP}
\frac{\mbox{d} \phi_c}{\mbox{d} t} = \mathcal{A}(k,t)\phi_c, ~~~ \phi_c(\xi,t_i) = \phi_c^0(\xi), ~~~ -\infty < \xi < \infty,
\end{equation}
where $t_i$ corresponds to the initial time, when perturbations are introduced and $\mathcal{A}(k,t)$ is the time dependent matrix. 

For the given  matrix $\mathcal{A}(k,t)$, we would like to have some effective way to determine whether one should be concerned about the effects of non-normality. The simplest quantitative approach often used for characterising normality is, $\kappa(V) \equiv || V ||_2 ~ || V^{-1} ||_2$, the condition number of the eigenvector matrix $V$ associated with $\mathcal{A}(k,t)$ \citep{Golub2007}. Here $|| \cdot ||_2$ corresponds to the standard Euclidean norm. It can be shown that for a normal matrix $\mathcal{A}(k,t)$ the condition number $\kappa(V)$ is $1$. In order to quantify the potential transient growth of disturbances and non-normality of $\mathcal{A}(k,t)$ associated to the IVP \eqref{eq:IVP}, first we compute  $\kappa(V)$ and then the numerical abscissa and the spectral abscissa, denoted by, $\alpha(\mathcal{A})$ and $\eta(\mathcal{A})$, respectively, and are defined as,
\begin{eqnarray}
\label{eq:numerical_abscissa}
& &\alpha(\mathcal{A})\equiv \max\{\Re (\lambda(\mathcal{A})\},\\
& &\eta(\mathcal{A})\equiv \max\{ \lambda(\mathcal{A + A^{\rm T}})/2)\}. 
\end{eqnarray}
Here $\mathcal{A} = \mathcal{A}(k,t)$, $\lambda(\cdot)$ represents the eigenvalue of the respective matrices, $\Re(\cdot)$ denotes the real part  and $\mathcal{A^{\rm T}}$ denotes the transpose of the matrix $\mathcal{A}$. The numerical abscissa $\eta(\mathcal{A})$ measures the maximum possible instantaneous growth rate corresponding to any initial condition as $t \to 0$ \citep{Trefethen2005}. It is important to note that for a normal matrix $\alpha(\mathcal{A}) = \eta(\mathcal{A})$. The scalar measures $\lambda(\cdot)$ and $\kappa(\cdot)$ of non-normality of the matrix $\mathcal{A}(k,t)$ are computed using the MATLAB routines $\mathsf{eig}$ and $\mathsf{cond}$, respectively.

\begin{table}
\centering
\begin{tabular}{ccccccc}
~ &\multicolumn{3}{c}{(i)} ~~~~ &\multicolumn{3}{c}{(ii)} \\
~ &\multicolumn{3}{c}{$U = 0, \mu(c) = 1, \rho(c) = c$}   ~~~~ &\multicolumn{3}{c}{$U = 1, \mu(c) = e^{1-c}, \rho(c) = 1$} \\ 
~ &\multicolumn{3}{c}{(DF in viscosity matched fluids)}   ~~~~ &\multicolumn{3}{c}{(VF in density matched fluids)} \\ \\ 
$t_0$ ~~~~ &$\alpha(\mathcal{A})$   & $\eta(\mathcal{A})$ & $\kappa(V)$  ~~~~  & $\alpha(\mathcal{A})$  & $\eta(\mathcal{A})$ & $\kappa(V)$  \vspace{0.2cm} \\
$0.1$ ~~~~ & -4.9626  & -2.4719    & 3.2121e+24    ~~~~ & -4.9658  & -2.4708    & 3.9351e+24  \\
$0.5$ ~~~~ & -0.9652  & -0.4730    & 7.5720e+24     ~~~~ & -0.9677  & -0.4724     & 9.2730e+24 \\
$1$ ~~~~ & -0.4671    & -0.2243     & 4.7560e+25    ~~~~ & -0.4692  & -0.2239     & 4.6768e+24  \\
$5$ ~~~~ & -0.0738   & -0.0291     & 4.8664e+24    ~~~~ & -0.0750  & -0.0290      & 8.9276e+24 \\
$10$ ~~~~ & -0.0276  & -0.0069      & 2.6092e+25    ~~~~ & -0.0285  & -0.0070      & 6.4083e+24 \\
$20$ ~~~~ & -0.0069  & {\bf 0.0023}      & 1.8240e+25     ~~~~ & -0.0075  &{\bf 0.0022} & 9.5316e+24 \\
$30$ ~~~~ & -0.0013  & {\bf 0.0043}      & 3.6163e+25     ~~~~ & -0.0017  & {\bf 0.0043} & 2.5617e+25 \\
$50$ ~~~~ & {\bf 0.0020}  &{\bf 0.0050}      & 4.4731e+24    ~~~~ & {\bf 0.0018}  & {\bf 0.0050}  & 5.2046e+25 
\end{tabular}
\caption{For a given wave number $k = 0.1$, the variation of spectral abscissa $\alpha(\mathcal{A})$, numerical abscissa $\eta(\mathcal{A})$, and the condition number of eigenvector matrix $\kappa(V)$, at different frozen time $t_0$: (i) DF in viscosity matched fluids, (ii) VF in density matched fluids.}
\label{table:NM1}
\end{table}

\begin{table}
\centering
\begin{tabular}{ccccccc}
~ &\multicolumn{3}{c}{(i)} ~~~~ &\multicolumn{3}{c}{(ii)} \\
~ &\multicolumn{3}{c}{$U = 0, \mu(c) = e^{c}, \rho(c) = c$}   ~~~~ &\multicolumn{3}{c}{$U = 1, \mu(c) = e^{1-c}, \rho(c) = c$} \\
~ &\multicolumn{3}{c}{(More viscous fluid at the top)}   ~~~~ &\multicolumn{3}{c}{(Less viscous fluid at the top)} \\ \\ 
$t_0$ ~~~~ &$\alpha(\mathcal{A})$   & $\eta(\mathcal{A})$ & $\kappa(V)$  ~~~~  & $\alpha(\mathcal{A})$  & $\eta(\mathcal{A})$ & $\kappa(V)$  \vspace{0.2cm} \\
$0.1$ ~~~~ & -4.9841  & -2.4930    & 5.4772e+24    ~~~~ & -4.9399  & -2.4427    & 7.4439e+24  \\
$0.5$ ~~~~ & -0.9851  & -0.4927    & 5.4482e+24     ~~~~ & -0.9429  & -0.4457    & 1.2781e+25 \\
$1$ ~~~~ & -0.4859    & -0.2429     & 1.6465e+25    ~~~~ & -0.4452    & -0.1984     & 9.9258e+24  \\
$5$ ~~~~ & -0.0890  & -0.0441     & 2.4518e+25    ~~~~ & -0.0540   & -0.0074     & 3.1297e+24 \\
$10$ ~~~~ & -0.0450  & -0.0200      & 2.0467e+24    ~~~~ & -0.0094  & {\bf 0.0125}      & 3.3447e+25  \\
$20$ ~~~~ & -0.0183  & -0.0088      & 6.1171e+24     ~~~~ & {\bf 0.0093}  & {\bf 0.0193}      & 8.1408e+24 \\
$30$ ~~~~ & -0.0114  & -0.0055      & 5.5201e+24     ~~~~ & {\bf 0.0138}  & {\bf 0.0199}      & 1.0704e+25 \\
$50$ ~~~~ & -0.0066  & -0.0034      & 5.3213e+24    ~~~~ & {\bf 0.0154}  &{\bf 0.0187}      & 2.2918e+25 
\end{tabular}
\caption{For a given wave number $k=0.1$, the variation of spectral abscissa $\alpha(\mathcal{A})$, numerical abscissa $\eta(\mathcal{A})$, and the condition number of eigenvector matrix $\kappa(V)$, at different frozen time $t_0$: (i) the influence of a stable viscosity contrast in the absence of displacement, (ii) the influence of an unstable viscosity contrast in the presence of displacement. }
\label{table:NM2}
\end{table}

The effect of non-normality in terms of the condition number, numerical abscissa and spectral abscissa for various flow conditions are summarized in tables \ref{table:NM1} and \ref{table:NM2}. The time dependent matrix $\mathcal{{A}}$ is frozen at different time and the computed $\alpha(\mathcal{A}), \eta(\mathcal{A})$ and $\kappa(V)$ are tabulated in table \ref{table:NM1} for classical VF of neutrally buoyant fluids and DF of viscosity matched fluids for a given wave number $k = 0.1$. For the case of VF the displacement velocity is taken as the characteristic velocity, such that $U = 1$, and the log-mobility ratio is $R = 1$. Next, we consider the influence of viscosity contrast on buoyantly unstable miscible fluids for $U = 0$ as well as $U \neq 0$. Table \ref{table:NM2} compares between two cases: (i) $\mu(c) = e^{c}, \rho(c) = c, U = 0$ and (ii) $\mu(c) = e^{1-c}, \rho(c) = c, U = 1$. Since $\kappa(V)$ is very large for all the cases discussed here, a substantial non-modal growth of the disturbances at early time can be anticipated, and this is confirmed from the difference between $\alpha(\mathcal{A})$ and $\eta(\mathcal{A})$ during initial period. Table \ref{table:NM1} depicts that the order of non-normality is almost equal for these two cases. From table \ref{table:NM2} it is identified that for $U = 0$ both $\alpha(\mathcal{A})$ and $\eta(\mathcal{A})$ are negative, which signifies that in the presence of the viscosity contrast onset of instability is delayed. On the other hand, for $U \neq 0$ instability is enhanced. These results are consistent with our observation from LSA as well as DNS discussed in \S \ref{sec:RD}.  

\begin{figure}
\centering
(a) \hspace{2in} (b) \\
\includegraphics[width=2in, keepaspectratio=true, angle=0]{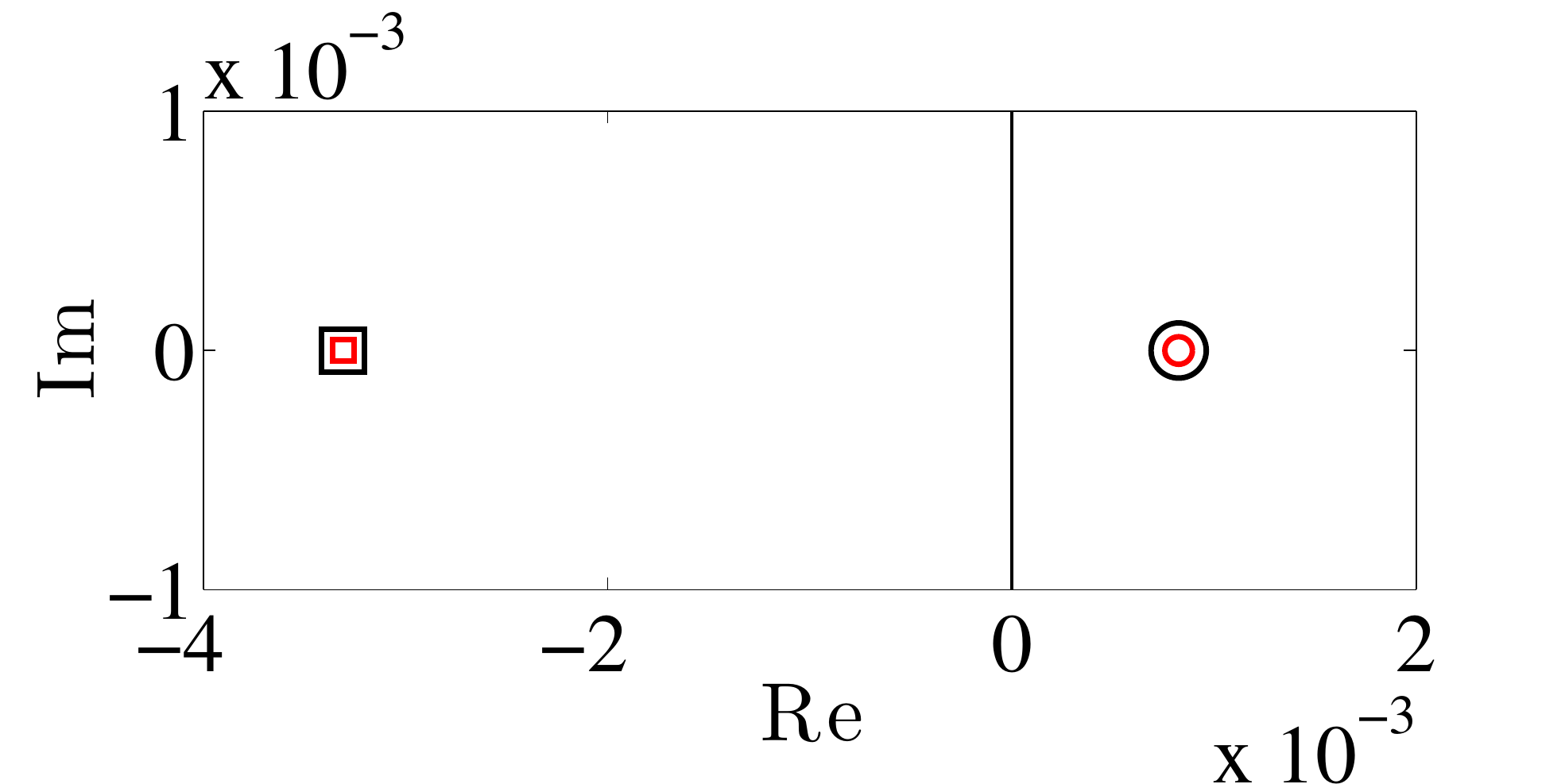}
\includegraphics[width=2in, keepaspectratio=true, angle=0]{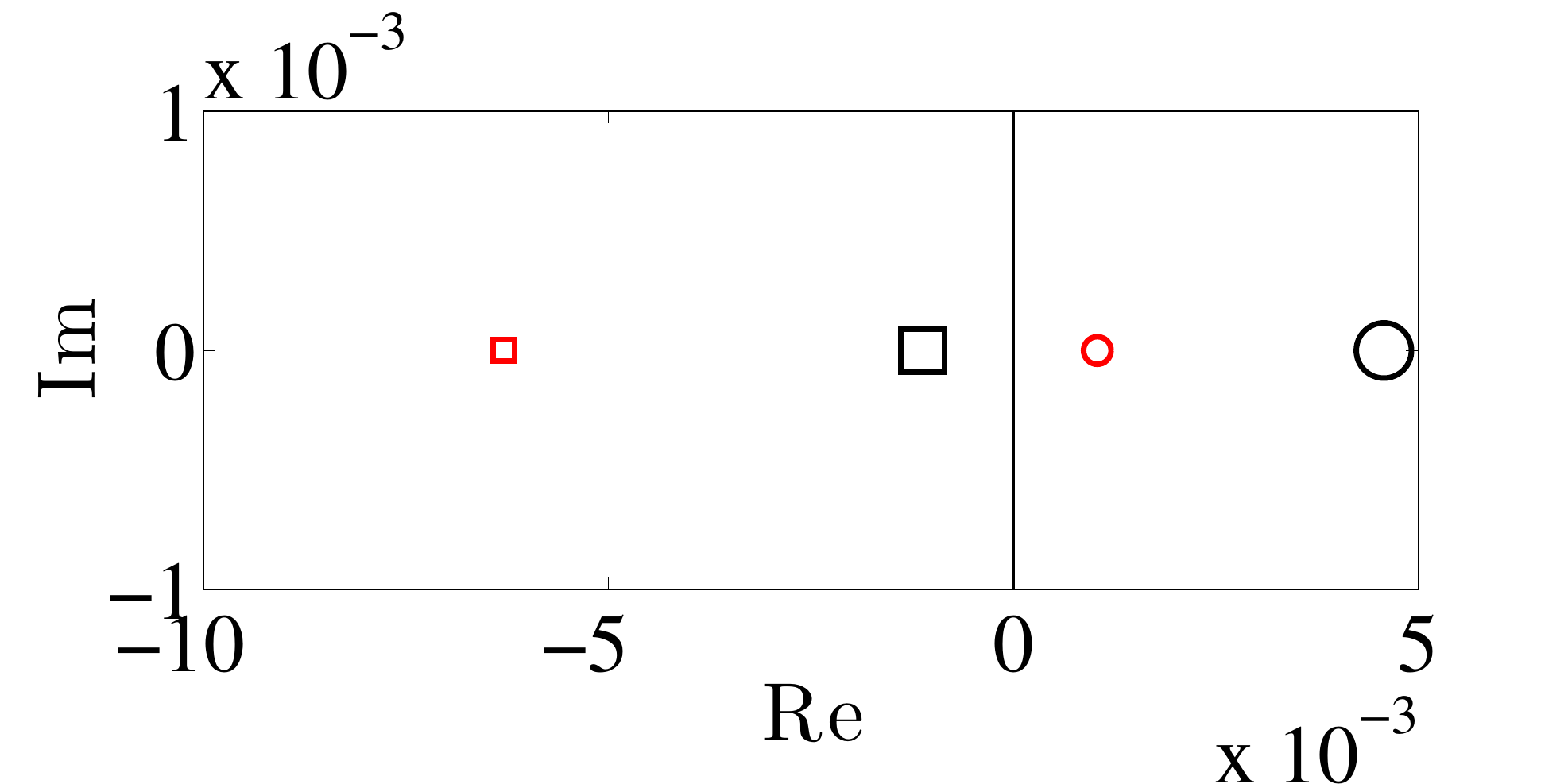}
\caption{Numerical abscissa ($\bigcirc$) and spectral abscissa ($\square$): (a) $k = 0.05, U = 0, R = 1, t_0 = 50$, (b) $k = 0.05, U = 0.5, R = 0.25, t_0 = 30$. Viscosity-concentration relation is colour coded; $\mu(c) = e^{Rc}$ (red) and $\mu(c) = e^{R(1-c)}$ (black). }\label{fig:NMA}
\end{figure}

Further, in order to understand the importance of appropriate viscosity scaling on the non-normal growth of the perturbations $\alpha(\mathcal{A})$ and $\eta(\mathcal{A})$ are plotted in figure \ref{fig:NMA}. Figure \ref{fig:NMA}(a) depicts that for a given wave number $k = 0.05$ and $U = 0$, $\alpha(\mathcal{A})$ and $\eta(\mathcal{A})$ corresponding to $\mu(c) = e^c$ are identical to those corresponding to $\mu(c) = e^{1-c}$. The influence of the viscosity contrast in the presence of fluid displacement (i.e. $U \neq 0$) on non-normality of the linearised matrix $\mathcal{A}$ is presented in figure \ref{fig:NMA}(b) for $U = 0.5, R = 0.25$ at $t_0 = 30$. This figure depicts that the numerical (spectral) abscissa corresponding to $\mu(c) = e^{R(1-c)}$ is larger than that corresponding to $\mu(c) = e^{Rc}$. Thus, we conclude that by choosing an appropriate characteristic viscosity one can lead to the same non-normal growth of the perturbations associated to the problem of a buoyantly unstable miscible interface both with the stable and unstable viscosity contrast when $U = 0$. On the other hand, in the presence of fluid displacement the instability is stronger when the less viscous fluid at the top displaces the more viscous fluid at the bottom. These observations are consistent with the LSA as well as the DNS presented in \S \ref{sec:RD} and could be captured only through the scaling analysis discussed in the present paper. 

To summarize, it is observed that the non-normality of the linearised matrix in the study of hydrodynamic instability driven by buoyancy or viscosity or both is of significant importance at early time. Although the frozen time approach to measuring the degree of non-normality of the time dependent matrix $\mathcal{A}$ through $\eta(\mathcal{A})$ and/or $\kappa(V)$ only provides a crude approximation, it can be handy to obtain an insight about possible non-modal growth of the disturbances. The transient growth of perturbations in a non-autonomous system can be determined efficiently through the propagator or matricant approach \citep{Rapaka2008, Hota2015a} or the direct adjoint looping (DAL) analysis \citep{Doumenc2010, Daniel2013}, which is beyond the scope of the current study. To determine the optimal perturbation leading to the instability in such cases is the topic of ongoing research and it is strongly believed that the importance of the scaling analysis discussed in this paper can also be observed in the optimal perturbations. 

\section{Conclusion}\label{sec:conclusion} 
We numerically investigate the influence of viscosity contrast on buoyantly unstable miscible interface in vertical porous media using an LSA as well as DNS. We show that in the absence of displacement a buoyantly unstable viscous miscible interface is the least stable when the viscosity of two fluids are equal, compared to the variable viscosity interface. In this case instability sets in at the same time for both less and more viscous upper fluid. On the other hand, a less viscous heavier fluid displacing a more viscous lighter fluid features an earlier onset than when the more viscous heavier fluid displaces the less viscous lighter fluid. We also show how a suitable rescaling of the dimensionless length, time and the displacement velocity can reproduce the results of Daniel and Riaz \cite{Daniel2014} from the present analysis and vice-versa. Thus the importance of an appropriate scaling analysis in fluid mechanics problems is presented by investigating the influence of viscosity contrast on buoyantly unstable miscible fluids in vertical porous media. The principal aim of an LSA is to obtain the onset of instability accurately and to predict the optimal perturbation that leads to the instability. Non-modal analysis to determine the optimal growth in buoyantly unstable miscible fluids with viscosity contrast has been undertaken for further study. 

\vspace{0.5in}
S.P. gratefully acknowledges the National Board for Higher Mathematics, Department of Atomic Energy, Government of India for the Ph.D. fellowship.

\end{document}